\documentstyle[preprint,amstex,amsfonts,amssymb,epsfig,aps]{revtex}
\tightenlines
\newcommand{\rv}{{\bf r}}

\newcommand{\sv}{{\bf s}}
\newcommand{\Rv}{{\bf R}}

\newcommand{\rd}{r_{12}}
\newcommand{\half}{{1\over 2}}

\renewcommand\eqref[1]{Eq.~(\ref{#1})}
\newcommand\klref[1]{(\ref{#1})}
\newcommand\cref[1]{Ref.~\cite{#1}}
\newcommand\abbref[1]{Fig.~\ref{#1}}
\newcommand\eqpref[1]{Eq.~(\protect\ref{#1})}
\newcommand\abbpref[1]{Fig.~\protect\ref{#1}}

\newcommand{\nn}{\nonumber}
\newcommand\av[1]{\left\langle #1 \right\rangle}
\newcommand\rhos{\rho^\ast}
\newcommand{\erf}{\operatorname{erf}}
\newcommand{\Tr}{\operatorname{Tr}}

\begin{document}
\draft
\title{Hard sphere-solids near close packing: Testing theories for crystallization}
\author{Benito Groh and Bela Mulder}
\address{
 FOM Institute for Atomic and Molecular Physics, Kruislaan 407, 1098
 SJ Amsterdam, The Netherlands}
\date{\today}

\maketitle

\begin{abstract}

The freezing transition of hard spheres has been well described by
various versions of density-functional theory (DFT).  These
theories should possess the close-packed crystal as a special limit,
which represents an extreme testing ground for the quality of such
liquid-state based theories. We therefore study the predictions of DFT
for the structure and thermodynamics of the hard-sphere crystal in
this limit. We examine the Ramakrishnan-Yussouff (RY)
approximation and two variants of the fundamental-measure theory (FMT)
developed by Rosenfeld and coworkers. We allow for general shapes of
the density peaks, going beyond the common Gaussian approximation. In
all cases we find that, upon approaching close packing, the peak width
vanishes proportionally to the free distance $a$ between the particles
and the free energy depends logarithmically on $a$. However, different
peak shapes and next-to-leading contributions to the free energy
result from the different approximate functionals. For the RY theory,
within the Gaussian approximation, we establish that the crystalline
solutions form a closed loop with a stable and an unstable branch both
connected to the close-packing point at $a=0$, consistent with the
absence of a liquid-solid spinodal. That version of FMT that has
previously been applied to freezing, predicts asymptotically step-like
density profiles confined to the cells of self-consistent cell
theory. But a recently suggested improved version which employs tensor
weighted densities yields wider and almost Gaussian peaks which are
shown to be in very good agreement with computer simulations.

\end{abstract}

\section{Introduction}

Some twenty years ago Alexander and McTague applied the formalism of
Landau theory to the freezing transition of atomic materials
\cite{Alexander:78}. Using symmetry arguments they suggested that a bcc
crystal should be the universally favored crystal structure,
independent of interaction details. This theory attempts to describe
the solid as a small, spatially periodic perturbation of a liquid. In
a recent paper \cite{Bif} we argued that such an approach should only
be valid near the liquid-solid spinodal, at which the liquid state
becomes locally unstable. The position of the spinodal is determined
by the Fourier transform of the liquid direct correlation
function $\tilde c$, and is given by the smallest density $\rho$ for which the equation 
\begin{equation} \label{spindef}
 \rho\, \tilde c(\rho,k)=1
\end{equation}
has a solution. Moreover, the perturbative approach does not apply to
the local minima of the free energy in order-parameter space, which
correspond to metastable or stable crystals, but rather to its saddle
points. For the latter we confirmed universal behavior near the
spinodal, which may have implications for nucleation \cite{Bif}. 

The
hard-sphere fluid has become the canonical model for freezing, since
it captures in the most simple form the dominant packing effects while
attractive interactions are believed to play only a secondary role.
The best current theories for hard-sphere freezing are various
versions of density-functional theory (DFT)
\cite{Singh:91,EvansRev:92,HaymetRev:92,Ohnesorge:94,AshcroftRev:95}.
Usually they are explicitely constructed to reproduce the
Percus-Yevick approximation $c_{PY}$ for the hard-sphere direct
correlation function. In \abbref{fig:cpspin} we show the values of
$\tilde c_{PY}(\rho,k)$ evaluated at the wave number $k_{max}(\rho)$
corresponding to the maximum at a given density $\rho$. One finds that
there is no solution to \eqref{spindef} at physical densities $\rho$
below the space filling density $6/\pi \sigma^{-3}$ where
$\sigma$ is the particle diameter (at and beyond this limit $c_{PY}$ is
not defined). This implies that those DFTs do not exhibit a
liquid-solid spinodal at all. Therefore the saddle point solution
branch of the stationarity equation derived from the density
functional cannot connect to the liquid branch when the bulk density
is increased. On the other hand, hard-core systems are characterized
by a close-packing density as the maximum possible density of a given
crystal structure. Upon approaching this limit a suitably defined
crystalline order parameter, e.g., the inverse width of the density
peaks, will diverge along the stable (minimum) branch. One may surmise
that that is also true along the saddle point branch. Thus an
alternative scenario to the bifurcation of a crystalline solution from
the liquid at a spinodal point as discussed in \cref{Bif}, are two
solid solution branches smoothly connected to each other at low
densities which diverge at close packing and are completely isolated
from the liquid. In order to test this hypothesis in the present work
we examine the close-packing limit in detail using DFTs that have
previously been applied to the low-density solid near the phase
transition.

Clearly, the strong localization of the particles in this limit
provides an extreme case for such liquid-state based theories. Hence
it is a good testing ground for assessing the qualities of different
approximations. In contrast to most DFT studies of the hard-sphere solid
we do not restrict the shape of the density peaks to Gaussians, but
allow for general spherically symmetric peaks. This is especially
interesting for the completely anharmonic hard-sphere crystal for
which there is no a priori argument to justify Gaussians, even for
small amplitude particle oscillations.

The starting point of density-functional theory is the free
energy functional of the inhomogeneous particle density $\rho(\rv)$
with the general form
\begin{equation}
  F[\rho(\rv)]=F_{id}[\rho(\rv)]+F_{ex}[\rho(\rv)].
\end{equation}
The ideal gas contribution is given by [$\beta=1/(k_B T)$]
\begin{equation}
  \beta F_{id}[\rho(\rv)]=\int d^3r \rho(\rv)[\ln\rho(\rv)\lambda^3-1]
\end{equation}
with the thermal de Broglie wavelength $\lambda$. While the excess
part $F_{ex}$ is not known exactly, a large number of approximate
forms have been suggested and applied to various problems in the last
decades \cite{Singh:91,EvansRev:92,HaymetRev:92,Ohnesorge:94,AshcroftRev:95}. As we do not strive for completeness we will
consider only two representative variants in this paper: the
Ramakrishnan-Yussouff functional \cite{Ramakrishnan:79,Haymet:83} which is one of the first and
simplest approximations that have been studied, and the fundamental
measure functional developed by Rosenfeld and coworkers \cite{Rosenfeld:89,Rosenfeld:97} which
at the present is believed to provide the best theoretical description
of the hard-sphere fluid. From a given functional the equilibrium
density distribution at a given bulk density $\rho_b$ is obtained by
minimization under the constraint $V^{-1} \int d^3
\rho(\rv)=\rho_b$. The value of the functional at its minimum is the
actual free energy of the system. For both functionals we performed
numerical calculations at a series of bulk densities as well as an
analytical analysis of the close-packing limit which enables us to
determine the asymptotic density profile and free energy.

\section{Ramakrishnan-Yussouff theory}

\subsection{Density functional and equilibrium profiles}

The Ramakrishnan-Yussouff functional follows from a density expansion
of $F_{ex}$ around the homogeneous state truncated at the quadratic
term:
\begin{equation} \label{FRY}
  \beta F_{ex}/V=\beta f_{ex}(\rho_b)-\frac{1}{2 V} \int d^3r d^3r'
  (\rho(\rv)-\rho_b) (\rho(\rv')-\rho_b) c(\bar\rho,|\rv-\rv'|).
\end{equation}
Here $f_{ex}$ is the free energy density and $c$ the direct
correlation function (DCF) of the hard-sphere liquid at an effective
density $\bar\rho$, both of which
are commonly approximated by the analytically known solutions of the
Percus-Yevick integral equation. In a
solid the density consists of a sum of identical peaks centered at the
lattice sites $\Rv$:
\begin{equation} \label{rhogen}
  \rho(\rv)=\sum_\Rv \rho_\Delta(\rv-\Rv).
\end{equation}
Throughout this paper it is assumed that the peaks are normalized
\begin{equation} \label{rhonorm}
  \int d^3r \rho_\Delta(\rv)=1
\end{equation}
and that the nearest-neighbour distance $R_{nn}$ in the lattice is
determined by the bulk density,
$R_{nn}/\sigma=(\rho_{cp}/\rho_b)^{1/3}$ where $\sigma$ is the particle
diameter and $\rho_{cp}$ is the maximum possible density. In order to
reduce the dimensionality of the integrations we moreover assume that
$\rho_\Delta$ is spherically symmetric. Deviations from this symmetry
exist \cite{Ohnesorge:93,Laird:87}, but are small especially near
close packing \cite{Young:74}. However, in contrast to most solid phase
calculations which assume $\rho_\Delta$ to be Gaussian here we do not
restrict its shape.

By insertion of \eqref{rhogen} in \eqref{FRY} one obtains
\begin{equation} \label{FRYsph}
  \beta F_{ex}/V=\beta f_{ex}(\rho_b)+\half \rho_b^2 \tilde
  c(\bar\rho,k=0)-\half \rho_b \sum_\Rv \int dr r^2
  \int dr' {r'}^2 \rho_\Delta(r) \rho_\Delta(r') w(r,r',R)
\end{equation}
where $\tilde c$ is the Fourier transformed DCF and the integral
kernel is given by
\begin{eqnarray} \label{wdef}
  w(r,r',R)&=&2\pi \int_0^{2\pi} d\phi_{12} \int_{-1}^1 d\cos\theta
  \int_{-1}^1 d\cos\theta' \nn \\
 & & {}\times c(\bar\rho,(r^2+{r'}^2+R^2+2 r R\cos\theta-2
  r' R \cos\theta'- 2 r r' \cos\gamma)^{1/2}).
\end{eqnarray}
The angles $\theta$, $\theta'$, and $\gamma$ are those between $\rv$
and $\Rv$, $\rv'$ and $\Rv$, and $\rv$ and $\rv'$, respectively, and
$\cos\gamma=\cos\theta \cos\theta'+\cos\phi_{12} \sin\theta
\sin\theta'$. The contribution from $R=0$ simplifies to
\begin{equation} \label{wR0}
  w(r,r',0)=\frac{8\pi^2}{r r'} \int_{|r-r'|}^{r+r'} d\rd \rd
  c(\bar\rho,\rd).
\end{equation}
Without loss of generality one may restrict the domain of
$\rho_\Delta$ to the Wigner-Seitz cell, so that the ideal contribution
to the functional can be written as
\begin{equation} \label{Fid}
  \beta F_{id}/V=4\pi \rho_b \int dr r^2
  \rho_\Delta(r)[\ln\rho_\Delta(r)\lambda^3 -1].
\end{equation}
By minimizing and taking into account the normalization
\eqref{rhonorm} one finds the stationarity equation
\begin{equation} \label{ELGgen}
  \rho_\Delta(r)=\frac{\exp[\frac{1}{4\pi} \sum_\Rv \int dr'
  {r'}^2 \rho_\Delta(r') w(r,r',R)]}{4 \pi \int dr r^2 \exp[\frac{1}{4\pi} \sum_\Rv \int dr'
  {r'}^2 \rho_\Delta(r') w(r,r',R)]}.
\end{equation}

The Percus-Yevick approximation for the hard-sphere DCF has the simple
form
\begin{equation}
  c(\bar\rho,r)=\left(c_0(\bar\rho)+c_1(\bar\rho) r+c_3(\bar\rho) r^3
  \right) \Theta(\sigma-r).
\end{equation}
The density-dependence of the coefficients $c_i$ can for example be
found in \cref{Hansen:86}. In the present context its most important
feature is the cutoff at the particle diameter which leads to
$w(r,r',R)=0$ for $R-r-r'>\sigma$. Hence for the strongly peaked
profiles in high density solids only the first shell of lattice
vectors ($|\Rv|=R_{nn}$) and the term with $R=0$ must be taken into
account. We have calculated $w(r,r',R_{nn})$ by numerical integration
using the trapezoidal rule with $50^3$ mesh points, while an
analytical expression for $w(r,r',0)$ was derived from
\eqref{wR0}. The stationarity equation is then discretized in $r$ and
solved by iteration. An underrelaxation scheme
\begin{equation}
  \rho^{(n+1)}=\omega \rho_{new}^{(n)}+(1-\omega) \rho^{(n)}
\end{equation}
proved helpful to ensure convergence. Here $\rho^{(n)}$ is the profile
after the $n$-th iteration and $\rho_{new}^{(n)}$ is the right hand
side of \eqref{ELGgen} calculated from $\rho^{(n)}$. A typical value
of the constant $\omega$ was 0.2.

The resulting profiles are shown in \abbref{fig:rhory}. Their width
scales with the free distance
$ a=R_{nn}-\sigma$
that a sphere can move into the direction to its neighbour if the
latter is kept fixed. The profile shapes approach a limiting form
discussed below. Their most striking property is the occurrence of a
maximum at intermediate distances $r$. This unphysical behavior
vanishes in the close-packing limit. The DCF has been evaluated at the
bulk density, $\bar\rho=\rho_b$. This most obvious choice has the
disadvantage that the solid has a higher free energy than the liquid
at all densities, as already pointed out in \cref{Baus:85}. In the
earliest DFT work the density of the coexisting liquid has been used
instead, but that is not very reasonable when high density solids are
considered. Other schemes to select a density $\bar\rho$
of an ``effective liquid'' have been proposed \cref{Baus:85,Baus:89,EvansRev:92}, which always imply
$\bar\rho<\rho_b$. Figure~\ref{fig:rhoryliq} shows density profiles
obtained with an arbitrarily chosen value
$\bar\rhos=\bar\rho\sigma^3=0.95$  which is close to the freezing
density. Now the maximum does not occur and the convergence to the
limiting shape is faster. The profiles are considerably flatter at
small $r$ than a Gaussian of the same width.

\subsection{Close-packing limit}

The results shown in Figs.~\ref{fig:rhory} and \ref{fig:rhoryliq}
clearly demonstrate that, in spite of contrary claims
\cite{Rosenfeld:96,Rosenfeld:98}, simple density-functional theories
based on the Percus-Yevick DCF do exhibit a well-defined close-packing
limit at which the peak width goes to zero. We will analyze this limit
in more detail in the following. Let us assume that for small
$a=R_{nn}-\sigma$ the profile behaves as
\begin{equation} \label{rhoasy}
  \rho_\Delta(r)=\frac{1}{\Delta^3}
  \rho_0\left(\frac{r}{\Delta}\right)
\end{equation}
with a width $\Delta=a/\alpha$ where $\Delta,a\to 0$ with $\alpha$
fixed. We shall show that the stationarity equation has a solution
consistent with these assumptions. The ideal free energy in this limit
becomes (with $N=\rho_b V$ and $s=r/\Delta$)
\begin{equation} \label{Fidasy}
  \beta F_{id}/N=4\pi \int_0^\infty ds s^2 \rho_0(s)
  \left[\ln\rho_0(s)-3\ln(\Delta/\lambda)-1\right].
\end{equation}
The relevant contributions to $F_{ex}$ are
\begin{equation}
  w(r=s\Delta,r'=s'\Delta,0)=16\pi^2 c(\bar\rho,0)+O(\Delta)
\end{equation}
and
\begin{eqnarray}
 w(r=s\Delta,r'=s'\Delta,R_{nn})&=&2 \pi \int_0^{2\pi} d\phi_{12}
 \int_{-1}^1 dx  \int_{-1}^1 dx'
 c(\bar\rho,\sigma[1+\Delta/\sigma(\alpha+sx-s'x')+
 O(\Delta^2)]) \nn \\
 & =&4\pi^2 c(\bar\rho,\sigma) \tilde w(s,s',\alpha)+O(\Delta)
\end{eqnarray}
where
\begin{eqnarray}
 \tilde w(s,s',\alpha)&=&\frac{1}{s s'} \int_{-s}^s ds_3 \int_{-s'}^{s'}
 ds'_3 \Theta(s'_3-s_3-\alpha) \nn\\
 &=& \left\{
 \begin{array}{ll}
  0, & s'+s<\alpha \\
  (s+s'-\alpha)^2/(2 s s'), & s'+s>\alpha, -\alpha<s'-s<\alpha \\
  2 (1-\alpha/s'), & s'-s>\alpha \\
  2 (1-\alpha/s), & s'-s<-\alpha 
 \end{array} \right.
\end{eqnarray}
Thus we finally have in leading order in $\Delta$
\begin{equation} \label{Fexasy}
  \beta F_{ex}/N=-2\pi^2 N_{nn} c(\bar\rho,\sigma) \int_0^\infty ds
  s^2 \int_0^\infty ds' {s'}^2 \rho_0(s) \rho_0(s') \tilde
  w(s,s',\alpha)+\mbox{const}=\Phi+\mbox{const}
\end{equation}
where $N_{nn}$ denotes the number of nearest neighbours.

The total free energy can now be minimized in two different
ways. First, one can restrict to profiles of a fixed shape
$\rho_0(s)$, e.g., Gaussians, and differentiate only with respect to
the scaled width $\alpha$ for fixed $a$ which gives
\begin{equation} \label{ELGal}
  3=-\alpha \frac{\partial\Phi}{\partial\alpha} 
\end{equation}
Due to the form of $w(s,s',\alpha)$ for $\alpha\to\infty$ one has
$\Phi\to 0$ and thus the right hand side of \eqref{ELGal}
also decays. On the other hand, for $\alpha\to 0$ $\Phi$ tends to a
positive constant (since $c(\bar\rho,\sigma)$ is negative), thus its
derivative will be negative for sufficiently well behaved
$\rho_0(s)$. Therefore the right hand side of \eqref{ELGal} is zero
both at $\alpha=0$ and $\alpha=\infty$ and positive in between
which implies a maximum at a finite value of $\alpha$. This can be
explicitely checked for Gaussians ($\rho_0(s)=\pi^{-3/2} \exp(-s^2)$)
and step functions ($\rho_0(s)=3/(4\pi) \Theta(1-s)$) for which the
integrals in \eqref{Fexasy} yield $\half
(1-\erf(\alpha/\sqrt{2}))$ and
$\half-\frac{3}{5}\alpha+\frac{1}{4}\alpha^3-\frac{3}{32}\alpha^4 
+\frac{1}{320}\alpha^5$. Depending on the height of this maximum
\eqref{ELGal} has zero or two solutions. In the first case there are
no stationary points with vanishing peak width at
$\rho_b=\rho_{cp}$. This is the case for the ``Onsager solid''
discussed in \cref{Bif} which belongs to the same class of approximate
functionals, but with $c(\bar\rho,r)$ replaced by its low-density
limit $-\Theta(\sigma-r)$. If $-c(\bar\rho,\sigma)$ is larger (e.g.,
$c_{PY}(\rho^{fcc}_{cp},\sigma)=-20.345$) the solution with smaller
$\alpha$ corresponds to a saddle point and the solution with larger
$\alpha$ to the stable solid minimum. We emphasize that the widths
$\Delta=a/\alpha$ for both solutions tend to zero for
$\rho_b\to\rho_{cp}$. In \abbref{fig:biflogges} we display the results
obtained for fcc and bcc solids, employing Gaussian profiles and
$\bar\rho=\rho_b$ (fcc: $\rhos_{cp}=\sqrt{2}$, $N_{nn}=12$; bcc:
$\rhos_{cp}=3\sqrt{3}/4$, $N_{nn}=8$). We also include numerical
solutions of $\partial F/\partial\Delta=0$ for the nonasymptotic
functional discussed above, evaluated for Gaussians. They approach the
asymptotics quite slowly, especially for the saddle points. At low
densities both branches are connected at an inflection point below
which no solidlike solutions exist.

Alternatively one can differentiate the asymptotic functional in
Eqs.~\klref{Fidasy} and \klref{Fexasy} with respect to the profile
$\rho_0(s)$. Here one may set $\alpha=1$ without loss of
generality. This leads to the Euler Lagrange equation
\begin{equation} \label{ELGasy}
  \rho_0(s)=\frac{\exp[\pi N_{nn} c(\bar\rho,\sigma) \int_0^\infty ds'
  {s'}^2 \rho_0(s') \tilde w(s,s',1)]} {4\pi \int_0^\infty ds\, s^2
  \exp[\pi N_{nn}  c(\bar\rho,\sigma) \int_0^\infty ds'
  {s'}^2 \rho_0(s') \tilde w(s,s',1)]}.
\end{equation}
Its solutions, which represent the asymptotic profile {\it shape},
obviously only depend on the value of $c$ at $r=\sigma$, because near
close packing the distance between two interacting particles is always
very close to $\sigma$. The resulting shapes, shown  in
Figs.~\ref{fig:rhory} and \ref{fig:rhoryliq}, are rather flat close to
the lattice site and decay strongly around $r/a=0.6$, so they
are definitely non-Gaussian.

The iteration never converged to a second solution that would repesent
the saddle point, even when started from the Gaussian saddle point
discussed above. It has been conjectured in a DFT study of the
isotropic-nematic transition of hard rods \cite{Kayser:78} that in
general the saddle point is not accessible by iteration because it corresponds to an unstable fixed point (see also \cref{Scheurle:77}).

We mention a subtle point in connection with \eqref{ELGasy}. Due to
the form of $\tilde w$ the right hand side goes to a constant for
$s\gg 1$, which means that no normalized solution on $[0,\infty)$ can
exist. However, as mentioned above, one may restrict to functions with
a finite support (e.g., $r<R_{nn}/2$, i.e., $s<R_{nn}/(2a)$). For the
numerical program indeed a much lower cutoff was used. In principle
the solution now depends on the cutoff, but in practice this
dependence is extremely weak because the constant approached for large
$s$ is of the order of $\exp(\frac{1}{2} N_{nn}
c(\bar\rho,\sigma))\simeq 10^{-53}$ so that the contributions from the
tail of $\rho_0(s)$ are neglegible for any reasonable value of the
cutoff. Similar remarks apply to \eqref{ELGgen}.

The free energy of the solid is determined by inserting the calculated
equilibrium profiles into the density functional. Its asymptotic
behaviour is given by
\begin{equation} \label{Ftotasy}
  \beta F/N=-3 \ln a+f_0+O(a).
\end{equation}
The leading logarithmic contribution stems from $F_{id}$ and is in
accordance with the result of free volume theory \cite{Buehler:50} and
cell theory \cite{Kirkwood:50,Wood:52}. It has been proven exact for
parallel hard cubes \cite{Hoover:65} and for finite hard-sphere
systems \cite{Salsburg:62} and 
is generally believed to be exact also in the thermodynamic limit. The various theories differ in
their prediction for the constant $f_0$. In the Ramakrishnan-Yussouff
approach (with $\bar\rho=\rho_b$) for an fcc solid we obtain
$f_0=21.7$ which is far above the molecular dynamics result
$f_0=-1.493$ \cite{Alder:68}. As shown in \abbref{fig:ftotlog} the
asymptotic form is approached quite slowly, i.e., the higher order
terms in \eqref{Ftotasy} are important up to high densities (which
probably will also produce a bad equation of state). The free energies
from the full minimization are only slightly below those for the best
Gaussian profile (\abbref{fig:ftotlog}).

\section{Fundamental-measure theory}

\subsection{Density functional} \label{DFTFMT}

Fundamental measure theory at present represents the best available
DFT for strongly inhomogeneous hard-sphere fluids. In contrast to most
previous approaches it does not depend on the direct correlation 
function as an input, but rather reproduces the Percus-Yevick
correlation function as an output of the theory in the homogeneous
limit.  While the originial expressions \cite{Rosenfeld:89} gave a
divergent excess free energy for strongly localized particles, a
recent empirical modification proved suitable also for the description
of the freezing transition \cite{Rosenfeld:97}. We will call this
version FMT1. Another new approximation has recently been suggested by
Tarazona and Rosenfeld \cite{Tarazona:97} based on more fundamental
grounds.  They presented a new derivation of FMT by enforcing the
functional to reduce to exactly known expressions in the zero- and
one-dimensional limit. They obtained a more complicated expression for
one of the excess free energy contributions that cannot be expressed
in terms of weighted densities and also does not reduce to the
Percus-Yevick free energy in the homogeneous limit. They also
suggested a simplification by rescaling a certain expansion of this
exact expression, which we adopt as FMT2. Due to its construction we
expect FMT2 to provide a better description of the high-density
crystal in which the individual particles are confined to quasi
zero-dimensional cages formed by their neighbors.

For a one-component hard-sphere fluid in three dimensions the
fundamental-measure functional has the form 
\begin{equation} \label{FexFMT}
\beta F_{ex}[\rho(\rv)]=\int d^3 r \sum_{i=1}^3 \phi_i(n_\alpha(\rv))
\end{equation}
where the functions $\phi_i$ depend only on the weighted densities
\begin{equation} \label{wddef}
  n_\alpha(\rv)=\int d^3r' \rho(\rv) w_\alpha(\rv-\rv')
\end{equation}
In FMT1 only two independent scalar and one vectorial
weight functions occur:
\begin{equation} 
  w_3(r)=\Theta(\frac{\sigma}{2}-r), \qquad
  w_2(r)=\delta(\frac{\sigma}{2}-r), \qquad
  {\bf w}_{V2}(\rv)=\frac{\rv}{r} \delta(\frac{\sigma}{2}-r),
\end{equation}
for FMT2 a tensor weight function is necessary:
\begin{equation}
 \hat{\bf w}_{ij}(\rv)=\frac{r_i r_j}{r^2} \delta(\frac{\sigma}{2}-r)
\end{equation}
The expressions for the excess free energy density are:
\begin{eqnarray} \label{phi1}
  \phi_1 & = & -\frac{n_2}{\pi\sigma^2} \ln(1-n_3), \\
  \label{phi2}
  \phi_2 & = & \frac{n_2^2-n_{V2}^2}{2\pi\sigma (1-n_3)}, \\
   \label{phi3}
  \phi_3^{FMT1} & = & \frac{(n_2^2-n_{V2}^2)^3}{24\pi n_2^3
  (1-n_3)^2}, \\
  \label{phi3FMT2}
  \phi_3^{FMT2} &=&\frac{9}{8\pi} \frac{\det\hat{\bf n}}{(1-n_3)^2}
\end{eqnarray}
The density ansatz \eqref{rhogen}  induces a corresponding form for
the weighted densities:
\begin{equation} \label{nanda}
  n_\alpha(\rv)=\sum_\Rv n_\Delta^{(\alpha)}(\rv-\Rv)
\end{equation}
with
\begin{equation}
  n_\Delta^{(\alpha)}(\rv)=\int d^3 r' \rho_\Delta(\rv')
  w_\alpha(\rv-\rv').
\end{equation}
If $\rho_\Delta$ is spherically symmetric the calculation of the
weighted densities reduces to one-dimensional integrations:
\begin{eqnarray}
 \label{n3sph}
  n_\Delta^{(3)}(r)&=&\frac{\pi}{r} \int_{|r-\sigma/2|}^{r+\sigma/2} dr'
  r' \left(\frac{\sigma^2}{4}-(r-r')^2\right) \rho_\Delta(r')
  +\Theta\left(\frac{\sigma}{2}-r\right) 4\pi \int_0^{\sigma/2-r} dr' {r'}^2
  \rho_\Delta(r') \\
  n_\Delta^{(2)}(r)&=& \frac{\pi\sigma}{r}
  \int_{|r-\sigma/2|}^{r+\sigma/2} dr' r' \rho_\Delta(r') \\
  {\bf n}_\Delta^{(V2)}(\rv)&=&\frac{\rv}{r}\frac{\pi}{r^2}
  \int_{|r-\sigma/2|}^{r+\sigma/2} dr' r'
  \left(r^2-{r'}^2+\frac{\sigma^2}{4}\right) \rho_\Delta(r').
\end{eqnarray}
In this case the
matrix $\hat{\bf n}_\Delta(\rv)$ [defined by $\hat{\bf n}(\rv)=\sum_\Rv
\hat{\bf n}_\Delta(\rv-\Rv)$] is diagonal in any coordinate system
aligned with $\rv$. An explicit calculation yields the eigenvalues
\begin{align} \label{n11}
  n^{(11)}_\Delta(r) & = n^{(22)}_\Delta(r)=\frac{\pi}{2r^3}
  \int_{|\sigma/2-r|}^{r+\sigma/2} dr' r' \left(4 r^2 {r'}^2 -
  (\frac{\sigma^2}{4}-{r'}^2-r^2)^2\right) \rho_\Delta(r') \\
\intertext{and} \label{n33}
  n^{(33)}_\Delta(r) & = \frac{\pi}{r^3}
  \int_{|\sigma/2-r|}^{r+\sigma/2} dr' r' 
  \left(\frac{\sigma^2}{4}-{r'}^2+r^2\right)^2 \rho_\Delta(r').
\end{align}
Note that $\Tr \hat{\bf n}_\Delta(\rv)=n_\Delta^{(2)}(r)$.
As $\rho_\Delta$ is a strongly peaked function of width $\Delta$ the
weighted densities $n_\Delta^{(2)}(\rv)$, ${\bf
n}_\Delta^{(V2)}(\rv)$, and $\hat{\bf n}_\Delta(\rv)$ have appreciable
values only for $|r-\sigma/2|\lesssim\Delta$ while $n_\Delta^{(3)}(r)$ tends to 1 for
much smaller $r$ and to 0 for much larger $r$. Thus for small $\Delta$
at any point $\rv$ in a solid at most two terms contribute appreciably
to the sum in \eqref{nanda}.

We only consider fcc solids. By exploiting the crystal symmetry the
integration in \eqref{FexFMT} can be restricted to a simplex
corresponding to 1/48 of the unit cell. In a coordinate
system aligned with the conventional cubic unit cell its vertices are
\begin{equation}
  (0,0,0), \qquad
  R_{nn} (\frac{1}{\sqrt{2}},0,0), \qquad
   R_{nn} (\frac{1}{2\sqrt{2}},\frac{1}{2\sqrt{2}},0), \qquad
   R_{nn} (\frac{1}{2\sqrt{2}},
  \frac{1}{2\sqrt{2}},\frac{1}{2\sqrt{2}}).
\end{equation}
It will be helpful to distinguish between the region A, that is
``affected'' by only one lattice site, and the region B affected by
two sites, i.e. the set of those points whose distance to two sites
differs from $\sigma/2$ by a length of order $\Delta$. As depicted in
\abbref{fig:regionab} region B consists of lens shaped sets around the
midpoints between neighboring sites. Here the integrands
$\phi_i(n_\alpha(\rv))$ do not depend on the azimuthal angle around
the line joining the sites, thus only a two-dimensional numerical
integration over cylindrical coordinates $\rho'$ and $z'$ must be performed. In order to
compute the full $\hat{\bf n}$ in region B the contribution from one
of the sites must be transformed to the coordinate system determined
by the direction to the other site. This is accomplished by a rotation
around an axis perpendicular to this direction by the angle $\gamma$
given by 
\begin{equation}
  \cos\gamma=\frac{\rv_+\cdot\rv_-}{r_+
  r_-}=\frac{{\rho'}^2+{z'}^2-R_{nn}^2/4}{{[({\rho'}^2+(z'+R_{nn}/2)^2)
 ({\rho'}^2+(z'-R_{nn}/2)^2)]}^{1/2}}
\end{equation}
Since in region A $n_\alpha(\rv)$
depends only on the distance to the nearest lattice site the
corresponding integration can even be reduced to one dimension after
the angular factors stemming from the shape of the simplex have been
worked out analytically. In practice
a sufficiently large cutoff $\Delta$ (typically $\Delta\simeq 2 a$)
was chosen beyond which $\rho_\Delta(r)$ is assumed to be zero, and
the integrals over A and B were calculated separately. This approach
proved to be much faster and more accurate than a straightforward 2d
integration over the whole simplex, because then the integrand is
essentially zero in large parts of the integration region.

\subsection{Equilibrium profiles} \label{DFTprof}

In order to determine the equilibrium density profile under the
constraint of spherical symmetry the functional derivatives of
$F_{ex}$ are calculated. We first write
\begin{equation} \label{dFex}
  \frac{\delta\beta F_{ex}}{\delta\rho_\Delta(r)}= \int d^3r'
  \sum_{i,\alpha} \frac{\partial\phi_i}{\partial n_{\alpha}}
  \frac{\delta n_{\alpha}(\rv')}{\delta \rho_\Delta(r)}
\end{equation}
and
\begin{equation} \label{dna}
  \frac{\delta n_{\alpha}(\rv')}{\delta \rho_\Delta(r)}=
  \frac{\delta}{\delta\rho_\Delta(r)} \sum_\Rv
  n_\Delta^{(\alpha)}(\rv'-\Rv) = \sum_\Rv \left.\frac{\delta
  n_\Delta^{(\alpha)}({\bf d})}{\delta\rho_\Delta(r)}\right|_{{\bf
  d}=\rv'-\Rv}.
\end{equation}
For $n_\Delta^{(3)}(d)$ the second term in \eqref{n3sph} is rewritten
as $\Theta(\sigma/2-d)(1-\int_{\sigma/2-d}^\infty dr' {r'}^2
\rho_\Delta(r'))$ which leads to
\begin{equation} \label{dn3d}
  \frac{\delta
  n_\Delta^{(3)}}{\delta\rho_\Delta(r)}=
  \left[\frac{\pi r}{d} \left(\frac{\sigma^2}{4}-(r-d)^2\right)-4\pi
  r^2 \Theta(\frac{\sigma}{2}-d)\right] \Theta(|d-\frac{\sigma}{2}|-r)
\end{equation}
where we assumed that always $r<\sigma/2+d$. Furthermore one finds
\begin{eqnarray} \label{dn2d}
  \frac{\delta
  n_\Delta^{(2)}(d)}{\delta\rho_\Delta(r)}&=&\frac{\pi\sigma r}{d}
  \Theta(|d-\frac{\sigma}{2}|-r) \\
  \label{dnv2d}
  \frac{\delta {\bf n}_\Delta^{(V2)}({\bf
  d})}{\delta\rho_\Delta(r)}&=&{\bf d} \frac{\pi r}{d^3}
  \left(d^2-r^2+\frac{\sigma^2}{4}\right) \Theta(|d-\frac{\sigma}{2}|-r).
\end{eqnarray}
For the tensor weighted density straightforward calculation leads to a
similar but more lengthy expression.
The partial derivatives $\partial\phi_i/\partial n_\alpha$ are easily
obtained from Eqs.~\klref{phi1}--\klref{phi3}. The functional
derivative can now be computed by inserting
Eqs.~\klref{dn3d}--\klref{dnv2d} into \eqref{dna} and that into
\eqref{dFex}. For the  integration over $\rv'$ in \eqref{dFex} we
adopt a similar scheme as for the
functional itself. Due to the step functions in
Eqs.~\klref{dn3d}--\klref{dnv2d} in region A the cutoff $\Delta$ can
be replaced by the distance
$r$ for which the derivative is evaluated. In region B two terms from
the lattice sum contribute. Because the integrand is nonanalytic at
the lines where one of the distances $d$ equals $\sigma/2-r$, $\sigma/2$, or
$\sigma/2+r$ we partitioned the integration region B appropriately for
the numerical integration. Together with the ideal free energy \eqref{Fid}
one readily obtains the stationarity equation
\begin{equation} \label{ELGFMT}
  \rho_\Delta(r)=\frac{\exp[-\frac{1}{4\pi r^2} \frac{\delta\beta
  F_{ex}/N}{\delta\rho_\Delta(r)}]}{4\pi\int dr' {r'}^2 \exp[-\frac{1}{4\pi {r'}^2} \frac{\delta\beta
  F_{ex}/N}{\delta\rho_\Delta(r')}]}.
\end{equation}
Again a mesh is introduced for $\rho_\Delta(r)$ and the weighted
densities are calculated by the trapezoidal rule with linear
interpolation between the mesh points. More sophisticated numerical integration
routines are used for the integration over
$\rv'$ in \eqref{dFex} for regions A and B, and \eqref{ELGFMT} is
iterated until the maximum relative change in $\rho_\Delta(r)$ is less
than $10^{-5}$.

The resulting profiles for FMT1 are displayed in
\abbref{fig:rhofmt}. They are almost constant at small $r$ and then
decrease steeper around $r=a/2$,
increasingly fast upon approaching the close-packing limit. In the
next section we show that the limiting shape indeed is a simple step
function. The profiles for FMT2 shown in \abbref{fig:rhofmtn} exhibit a much smoother,
Gaussian-like decay and their  width, measured, e.g., by
$\av{r^2}=\int d^3 r\, r^2 \rho_\Delta(r)$, on the scale $a$ is
considerably larger than for both the RY and FMT1
functionals. Clearly, again the absolute width goes to zero
linearly with $a$, as expected in the close-packing limit.

\subsection{Close-packing limit} \label{cpFMT}

As for the Ramakrishnan-Yussouff functional we assume that
asymptotically the density profile has the form \eqref{rhoasy} with
$\Delta=a=\sigma\delta$. We have seen that  in the important range the
argument of the weighted densities is close to $\sigma/2$. Therefore
we set $r/\sigma=1/2+t\delta$ and determine the leading contributions
to $n_\Delta^{(\alpha)}(r)$ for small $\delta$ and fixed $t$:
\begin{eqnarray} 
  n_\Delta^{(3)}(t) & = & \sum_{i=0}^\infty n_{3i}(t) \delta^i
   =2\pi\Bigg[2\Theta(-t) \int_0^{-t} ds s^2 \rho_0(s) \nn \\
  \label{n3lim}
   & &{}+\int_{|t|}^\infty ds s \rho_0(s)
   \left((s-t)+\delta (t^2-s^2)+2\delta^2 t (s^2-t^2)+\cdots\right)\Bigg]
   \\
  n_\Delta^{(2)}(t)&=&\frac{1}{\sigma}\sum_{i=0}^\infty n_{2i}(t)
   \delta^{i-1} \nn\\
  \label{n2lim}
   & = & \frac{2\pi}{\sigma\delta} \int_{|t|}^\infty ds s \rho_0(s)
   \left(1-2t\delta+4t^2 \delta^2+\cdots\right) \\
  \label{nv2lim}
  {\bf n}_\Delta^{(V2)}(\hat\rv,t) & = & \hat\rv
   \frac{2\pi}{\sigma\delta} \int_{|t|}^\infty ds s \rho_0(s)
   \left(1-2 t\delta+(6t^2-2s^2) \delta^2+\cdots\right) \\
  n_\Delta^{(11)}(t)&=&4\pi\delta \int_{|t|}^\infty ds s \rho_0(s)
   \left(s^2-t^2 +\cdots\right) \\
  n_\Delta^{(33)}(t) & = & \frac{2\pi}{\delta} \int_{|t|}^\infty ds s
   \rho_0(s) \left (1-2 t\delta +4 (2 t^2-s^2) \delta^2 +\cdots\right)
\end{eqnarray}
where the caret denotes a unit a vector. Since for any $\rho_0(s)$ the
first two terms in the expansions of $n_\Delta^{(2)}$ and $|{\bf
n}_\Delta^{(V2)}|$ are identical one has $n_2^2-n_{V2}^2=O(\delta^0)$
in region A. On the other hand, in region B the contributions to ${\bf
n}_{V2}$ from the two lattice sites have almost opposite directions so
that $n_2^2-n_{V2}^2\sim \delta^{-2}$ there. For FMT2 we find
$\det\hat{\bf n}\sim\delta$ in both regions, because, due to the
quadratic dependence of ${\bf w}_{ij}(\rv)$ on the components of
$\rv$, the two contributions do not cancel each other in region B.
Taking into account that the volumes of A and B are proportional to
$\delta$ and $\delta^2$, respectively, we can estimate the order of
the individual free energy contributions $\Phi_i=N^{-1} \int d^3 r
\phi_i(n_\alpha(\rv))$:

\bigskip
\begin{tabular}{lll} 
 & A & B \\
$\Phi_1$ & $\delta^0$  & $\delta  $ \\
$\Phi_2$ & $\delta  $  & $\delta^0$ \\
$\Phi_3^{FMT1}$ & $\delta^4$  & $\delta^{-1}$  \\
$\Phi_3^{FMT2}$ & $\delta^2$ & $\delta^3$
\end{tabular}
\bigskip

Thus at this point a qualitative difference between the two
approximations arises, as different terms become dominant in the
close-packing limit. We first discuss FMT1, for which $\Phi_{3B}$ is
the leading term. In a cylindrical coordinate system
$(z',\rho',\phi')$, centered at the midpoint between two sites and
with its axis directed towards (see \abbref{fig:regionab}) one of
them, the distances $r_\pm$ to the sites, which occur as the argument
of the weighted densities $n_\Delta^{(\alpha)}$, are
\begin{equation} \label{rpm}
  r_\pm=\left[{\rho'}^2+(z'\pm R_{nn}/2)^2\right]^{1/2}.
\end{equation}
In the scaled coordinates $\rho={\rho'}^2/(\delta\sigma^2)$ and
$z=z'/(\delta\sigma)$ one has $t_\pm=1/2+\rho\pm z+O(\delta)$ and 
\begin{equation}
  n_2^2-n_{V2}^2=\frac{4}{\delta^2} n_{20}(t_+) n_{20}(t_-)+\cdots
\end{equation}
which finally yields
\begin{equation} \label{Phi3Blim}
  \Phi_{3B}^{FMT1}\simeq\frac{32}{\delta} \int_0^\infty d\rho \int_0^\infty dz
  \left(\frac{n_{20}(t_+)
  n_{20}(t_-)}{n_{20}(t_+)+n_{20}(t_-)}\right)^3
  \frac{1}{(1-n_{30}(t_+)-n_{30}(t_-))^2}.
\end{equation}
Since $n_{30}(t)\in[0,1]$ and $n_{20}(t)\geq 0$ this expression is
positive. It attains its minimum value zero for all profiles
$\rho_0(s)$ that have a strict cutoff at $s=1/2$ so that  region B is
empty. In this restricted class of profiles the dominant contributions
are $\Phi_{1A}$ and $F_{id}$. The former can be written as 
\begin{equation}
  \Phi_{1A}=-\int_{-1/2}^{1/2} dt\, n_{20}(t) \ln(1-n_{30}(t))+O(\delta).
\end{equation}
But the fact that $n_{20}(t)=-\partial n_{30}/\partial t$ implies
$\Phi_{1A}=1+O(\delta)$ for all profiles. Since here the peaks around
different sites are independent of each other, this result is
consistent with the extensively discussed 0D limit of the
fundamental-measure functional \cite{Rosenfeld:97,Tarazona:97}: For density profiles
$\rho_\Delta(\rv)$ 
constrained to a volume that cannot hold more than one particle the
exact excess free energy is $\beta F_{ex}=1$ if $\int d^3 r
\rho_\Delta(\rv)=1$. One of the merits of the present theory is that
this limit is almost exactly fulfilled \cite{Rosenfeld:97}. At last we
are left with the ideal free energy \eqref{Fidasy}
as the only relevant $O(\delta^0)$ term, which, naturally, favors an
evenly distributed density:
\begin{equation}
  \rho_0(s) \to \frac{6}{\pi} \Theta(\half-s).
\end{equation}

This finding implies that the usually assumed Gaussian peaks represent
a particularly bad approximation in this case. Indeed, in the Appendix
we show that the width $\Delta$ of the best Gaussian is asymptotically
related to the free distance $a$ by $a\sim \Delta\sqrt{\ln(-\Delta)}$
which means that the ratio $\Delta/a$ tends to zero, albeit very
slowly. The intuitive reason is that the tail of the Gaussian profile
leads to an unfavorable free energy contribution $\Phi_{3B}$ that can
only be kept small if the tail increasingly ``retracts''.

Actually the above arguments for the asymptotic step function shape in
FMT1
can be generalized to nonspherical profiles. Starting from
\begin{equation}
  \rho_\Delta(\rv)=\frac{1}{\Delta^3} \rho_0(\rv/\Delta)
\end{equation}
and setting again $\sv=\rv/\Delta$, $\Delta=\sigma\delta=a$, and
$|\rv|/\sigma=1/2+t\delta$ one has
\begin{equation}
  n_\Delta^{(3)}(\hat\rv,t)=\int d^3 s\,
  \Theta(\hat\rv\sv-t-\delta(t^2-2 t \hat\rv\sv+s^2)) \rho_0(\sv).
\end{equation}
Expanding for small $\delta$ gives
\begin{equation}
 n_\Delta^{(3)}(\hat\rv,t)=\int d^3 s\,
  \Theta(\hat\rv\sv-t)\rho_0(\sv) +O(\delta).
\end{equation}
Analogously we find
\begin{equation}
 n_\Delta^{(2)}(\hat\rv,t)=\int d^3 s \rho_0(\sv)
 \left[\frac{1}{\delta} \delta(\hat\rv\sv-t)-(t^2-2 t \hat\rv\sv+s^2)
 \delta'(\hat\rv\sv-t)+O(\delta)\right]
\end{equation}
and, using ${\bf n}_\Delta^{(V2)}(\rv)=-\nabla n_\Delta^{(3)}(\rv)$,
\begin{eqnarray}
  n_\Delta^{(2)}(\hat\rv,t)&=&\int d^3 s \rho_0(\sv)
 \Bigg[\frac{1}{\delta}\hat\rv \delta(\hat\rv\sv-t)-
 \hat\rv(t^2-2 t \hat\rv\sv+s^2)
 \delta'(\hat\rv\sv-t) \nn \\
 & &{}-2 (\sv-(\hat\rv\sv)\hat\rv)\delta(\hat\rv\sv-t)+O(\delta)\Bigg]. 
\end{eqnarray}
Since the last term in this equation is perpendicular to $\hat\rv$ the
combination $n_2^2-n_{V2}^2$ is still of order $\delta^0$ in region
A. If in region B the same coordinates $(z,\rho,\phi')$ as before are
used and
the vectors to the nearest lattice sites are denoted by $\rv_{\pm}$, the
fact that $\hat\rv_+\hat\rv_-=-1+O(\delta)$ yields
$n_2^2-n_{V2}^2=O(\delta^{-2})$. Thus, in summary all estimates for
the individual terms given in the table above remain valid. Again the
dominant term $\Phi_{3B}$ is positive and minimized by cutoff profiles. As
the leading terms of the scalar weighted densities are related by
$n_{20}(\hat\rv,t)=-\partial n_{30}(\hat\rv,t)/\partial t$ the
contribution $\Phi_{1A}$
in leading order is still independent of the profile. The ideal term
now enforces $\rho_\Delta(\rv)$ to be constant in the maximum allowed
region C that is compatible with $B=\emptyset$. It can be constructed
by shifting the bounding planes of the Wigner-Seitz cell inward by
$\sigma/2$ (see \abbref{fig:cell}). A given point $\rv$ in C contributes to the weighted
densities at $\rv'$ only if $|\rv-\rv'|\leq\sigma/2$. By construction
all such $\rv'$ lie within the Wigner-Seitz cell and thus cannot be
``reached'' from any $\rv$ in the cell C' around another site, which
means that B is indeed empty. However if a point P outside of C were
added the distance to its mirror point P' with respect to the closest
Wigner-Seitz boundary plane would be less than $\sigma$ so that
elements of B would lie on their joining line (see
\abbref{fig:cell}). The cell C constructed here is
identical to that of the self-consistent cell theory
\cite{Kirkwood:50}. Its volume for an fcc solid is $a^3/\sqrt{2}$.

We now turn to the second approximation (FMT2) for which $\Phi_{2B}$
and $F_{id}$ are the dominant contributions. (Remember that $\Phi_{1A}$ is independent of
$\rho_0(s)$ in leading order.) Analogous to \eqref{Phi3Blim} we have,
up to higher orders in $\delta$,
\begin{equation} \label{Phi2Blim}
  \Phi_{2B}\simeq 24 \int_0^\infty d\rho \int_0^\infty dz
  \frac{n_{20}(t_+) n_{20}(t_-)}{1-n_{30}(t_+)-n_{30}(t_-)}
\end{equation}
with $t_\pm=1/2+\rho\pm z$. The corresponding stationarity equation is
\begin{equation} \label{ELGasyFMT2}
  \rho_0(s)=\frac{\exp\left(-\frac{1}{4\pi
  s^2}\frac{\delta\Phi_{2B}}{\delta\rho_0(s)}\right)} 
  {4\pi \int_0^\infty ds' {s'}\exp\left(-\frac{1}{4\pi
  {s'}^2}\frac{\delta\Phi_{2B}}{\delta\rho_0(s')}\right)}.
\end{equation}
The functional derivative is calculated as in Sect.~\ref{DFTprof}. The
resulting asymptotic profile shown in \abbref{fig:rhofmtn} is close to
those obtained for finite densities using the full functional. It is
almost, but not exactly Gaussian.

The asymptotic free energy of the fundamental-measure theory also has
the form \eqref{Ftotasy}. In FMT1 one has  $f_0=\half \ln 2=0.3466$.
Under the constraint of spherical symmetry this is replaced by
$f_0=\ln(6/\pi)=0.6470$, both of which are much closer to the correct
value than the Ramakrishnan-Yussouff theory. Results for finite
distance from close packing are given in \abbref{fig:ftotlog} and
agree, probably by accident, rather well with the computer
simulations. The approach to the
asymptotic law is quite slow. On the
other hand for FMT2 not only the profiles but also the
free energies  (\abbref{fig:ftotlog}) approach their asymptotic limit
faster in this version of the theory. The value of the constant in
\eqref{Ftotasy} is found to be $f_0=-1.527$ in very good agreement
with the MD results. However, in view of the relatively large change
in $f_0$ due to nonsphericity of the profiles as observed for FMT1,
this may well be fortuitous. We did not consider nonspherical profiles
in FMT2.

\subsection{Saddle point} \label{saddle}

In view of the discussion in the introduction it would be interesting
also to keep track of the saddle point between the liquid and the
solid state when close packing is approached. Unfortunately, again one
is plagued by the fact that the iteration of the stationarity equation
does not converge to a second solution. Furthermore the arguments of
the asymptotic analysis do not apply to the saddle point because they
essentially involve a minimization in two steps. Hence one must revert
to parametrizations of the density with a few parameters. For
Gaussians and step functions the saddle point occurs at a width
$\Delta$ proportional to $a^{2/3}$ in FMT1 and to $a^{1/2}$ in
FMT2. However, a priori there is no reason to assume that at the
saddle point the profile has a similar shape as at the minimum. We
also tried profiles of the forms $\rho_\Delta(r)\sim
\exp(-(r/\Delta)^n)$ and $\rho_\Delta(r)\sim (1+r/\Delta)^{-n}$ and
found numerically that in both cases the free energy at the maximum
with respect to $\Delta$ decreases with decreasing $n$, down to the
lowest feasible values of $n$. This suggests that the actual saddle
point profile may decay very slowly, while within these restricted
classes of profiles a true saddle point at a nondegenerate profile
seems not to exist.

\section{Monte Carlo simulations}

Although an extensive computer simulation study of the density
distribution in hard-sphere crystals has been carried out before
\cite{Young:74}, no useful results for the radial distribution
function have been published. In order to assess the quality of
the various theories we therefore undertook a small Monte Carlo (MC)
simulation ourselves. In an NVT ensemble of $8^3$ spheres in an fcc
arrangement we measured the distribution of the particles' distance
$r$ from their equilibrium sites. We corrected for the movement of
these sites due to shifts in the center of mass. Measurements were
taken over $2\cdot 10^6$ MC steps per particle for two bulk
densities. The results are plotted in \abbref{fig:rholog} on a
logarithmic scale versus $(r/a)^2$ and compared to the various DFT
calculations. The quantitative agreement is excellent for FMT2. The profiles
are close to Gaussians but decay faster at large distance than a
Gaussian fitted to the small distance part. The dependence of the
scaled profiles on bulk density is rather small in the examined range,
but still qualitatively reproduced by the theory. The actual width of
the profiles will increase with increasing particle number
\cite{Young:74}, but we did not attempt to correct for finite size
effects. In \cref{Young:74} it was found that in
the thermodynamic limit for high densities the width behaves as
$\av{r^2}^{1/2}/a=1.098\pm 0.004$, again in almost perfect agreement
with FMT2, for which $\av{r^2}^{1/2}/a=1.025$. This means that FMT2 is
the first DFT which yields the correct value of the Lindemann parameter.

\section{Summary and Discussion}

In summary, we have analyzed the close-packing limit of the hard-sphere crystal using three versions of DFT. All of them predict a peak
width $\Delta$ that vanishes proportional to the free distance $a$ and
yields a logarihmic term in the free energy (see \eqref{Ftotasy})
stemming from the ideal gas entropy. Numerically this has been
observed before, for Gaussian peaks, in two other DFTs, the
generalized liquid approximation (GELA) and the modified weighted
density approximation (MWDA) \cite{Tejero:95}. For the latter,
however, it was found later that the solutions correspond to
``unphysical'' branches \cite{Tejero:97}. The relative performance of
the different theories can be judged from the profile shape obtained
by free minimization. RY gives too narrow profiles with an unphysical
maximum if the bulk density is used as the expansion point
(\abbref{fig:rhory}). The shape and width are also wrong for other
expansion points (\abbref{fig:rhoryliq}). FMT1 predicts asymptotically
steplike profiles confined to the cells of cell theory
(\abbref{fig:rhofmt}). Only the FMT2 profiles (\abbref{fig:rhofmtn})
are in quantitative agreement with simulations at high densities
(\abbref{fig:rholog}). In spite of the anharmonicity of the hard-sphere crystal they are close to Gaussians. Similarly the results for
the next-to-leading free energy contribution improve from RY to FMT1
to FMT2 (\abbref{fig:ftotlog}), the two FMT versions being much closer
to the correct result than RY. This could have been expected from the
way the RY approximation is constructed: A density expansion around a
liquid state certainly is difficult to justify for the highly ordered
high-density crystal.

If one restricts the profiles to a fixed shape saddle points of the
free energy are found at widths decaying $\sim a^x$ with $x_{RY}=1$,
$x_{FMT1}=2/3$, and $x_{FMT2}=1/2$. Insofar the global scenario for
the crystalline solutions proposed in the introduction is comfirmed
(see also \abbref{fig:biflogges}). However, as detailed in
Sec.~\ref{saddle}, in larger classes of functions the saddle point
remains elusive. We remark that the saddle point is a property closely
connected to the mean-field type free energy functional and, e.g., is
not directly accessible by computer simulations.

Comparing the two variants of FMT we see that the structure in the
close-packing limit is sensitive to subtle differences between DFT
approximations and thus might be a guiding line in the construction of
better FMT-like functionals. Besides our FMT1 some other
approximations for $\phi_3$ have been suggested in \cref{Rosenfeld:97}
which are all of the form
\begin{equation}
  \phi_3=\frac{n_2^3}{(1-n_3)^2} f(\xi)
  \qquad\text{with}\qquad
  \xi=|\boldsymbol{\xi}|=\left|\frac{{\bf n}_{V2}}{n_2}\right|.
\end{equation}
The power of $n_2$ is determined by dimensional arguments and the
function $f$ can only depend on the absolute value of $\boldsymbol{\xi}$ because
of the isotropy of space. From Eqs.~\klref{n2lim} and \klref{nv2lim}
we have $\xi=1+O(\delta^2)$ in region A. In order not to spoil the
correct leading order for a quasi-zero-dimensional situation as given by
$\Phi_{1A}$ one has the additional requirement $f(\xi\to 1)\sim
(1-\xi)^n$ with $n\geq 2$, which implies $\Phi_{3A}\sim
\delta^{2n-2}$. But in region B $\xi$ varies between zero and one so
that always $\Phi_{3B}\sim \delta^{-1}$. The function $f$ must be
nonnegative in this range, otherwise the functional would not be
bounded from below [this happened in the original FMT
\cite{Rosenfeld:89} for which $f(\xi)=(1/3-\xi^2)/(8\pi)$]. Then the
argument of Sec.~\ref{cpFMT} runs through and the asymptotic profiles
will always be step functions. We conclude that an improved
description of the high density solid is not possible within FMT if
only the scalar and vector weighted densities are used, the tensor
weight function of FMT2 is inevitable. On the other hand in FMT2 the
behavior near close packing is exclusively determined by $\phi_2$ so
that  no  conditions on the precise form of $\phi_3$ can be deduced.

\section*{acknowledgments}

We thank Sander Pronk for providing the Monte Carlo program and
Y. Rosenfeld for stimulating discussions.
This work is part of the research program of the Stichting voor
Fundamenteel Onderzoek der Materie (Foundation for Fundamental Research on
Matter) and is made possible by financial support from the Nederlandse
Organisatie voor Wetenschappelijk Onderzoek (Netherlands Organization for
the Advancement of Research). B.G. acknowledges the financial support
of the EU through the Marie Curie TMR Fellowship programme.

\begin{appendix}

\section{Gaussian peaks in FMT1}

For Gaussian density peaks $\rho_0(s)=\pi^{-3/2} e^{-s^2}$ the leading
contributions to the weighted densities are (see
Eqs.~\klref{n3lim}--\klref{nv2lim})
\begin{equation}
  n_{30}(t)=\half(1-\erf(t)) \qquad
  n_{20}(t)=\frac{1}{\sqrt{\pi}} e^{-t^2}
\end{equation}
and for a width $\Delta=a/(2\alpha)$ the dominant excess free energy
contribution is (see \eqref{Phi3Blim})
\begin{multline}
  \Phi_{3B}=\frac{32}{\pi^{3/2}} \frac{\sigma}{\Delta} \int_0^\infty
  d\rho \int_0^\infty dz
  \left[e^{(\rho+\alpha+z)^2}+e^{(\rho+\alpha-z)^2}\right]^{-3} \\
  {}\times\left[\half\erf(\rho+\alpha+z)+\half\erf(\rho+\alpha-z)\right]^{-2}
  +O(\Delta^0).
\end{multline}
In order that $F_{ex}$ does not become too large for $a\to 0$ we
expect $\alpha\to\infty$. In this limit the substitutions
$\rho'=\rho\alpha$ and $z'=z\alpha$ yield
\begin{equation}
  \Phi_{3B}=\frac{1}{12\sqrt{\pi}} \frac{\sigma}{\Delta} \frac{\exp(-3
  \alpha^2)}{\alpha^2}.
\end{equation}
Now we can add 
\begin{equation}
  \beta F_{id}/N=-\frac{3}{2}\ln\pi(\Delta/\lambda)^2-\frac{5}{2}
\end{equation}
and minimize with respect to $\Delta$ which gives
\begin{equation}
  \frac{\sigma}{\Delta} \frac{1}{2\sqrt{\pi}}
  \exp(-\frac{3}{4}\frac{a^2}{\Delta^2}) =3.
\end{equation}
This equation indeed has a solution with $\Delta/a\to 0$ for $a\to 0$;
solved for $a$ one has
\begin{equation}
  a=\frac{2\Delta}{\sqrt{3}}
  \left[-\ln\left(6\sqrt{\pi}\frac{\Delta}{\sigma}\right)\right]^{1/2}
\end{equation}
which demonstrates that $\Delta/a$ decays only very
slowly. Nevertheless this decay is at variance with the physical
expectation $\Delta/a\to\text{const}$ which is well supported by
computer simulations \cite{Young:74}, and, as shown in the main text,
is also fulfilled within the present theory if allowance is made for
more general profile shapes.

\end{appendix}


\begin{figure}
\caption{The left-hand side of \eqpref{spindef} for the hard-sphere
direct correlation function in the Percus-Yevick approximation. The
wave number $k_{max}(\rho)$ corresponds to the maximum of
$c_{PY}(\rho,k)$ at a given density $\rho=\rho^\ast\sigma^{-3}$. The
curve lies below unity for all admissable densities
$\rho^\ast<6/\pi=1.910$, i.e., for packing fractions
$\eta=\rho^\ast\pi/6<1$, which means that there is no liquid-solid spinodal. The close-packing limit occurs at
$\rho^\ast=\protect\sqrt{2}$.}
\label{fig:cpspin}
\end{figure}

\begin{figure}
\caption{Density profiles in a  high density fcc crystal calculated
from Ramakrishnan-Yussouff DFT. Note that the distance $r$ from
the lattice site and the density are scaled by the free distance
$a=R_{nn}-\sigma$, which varies over 2.5 orders of magnitude in this
density range.}
\label{fig:rhory}
\end{figure}

\begin{figure}
\caption{The same as \abbpref{fig:rhory} but using $\bar\rhos=0.95$ as
the density argument of the DCF. In this case the profiles are monotonic.}
\label{fig:rhoryliq}
\end{figure}

\begin{figure}
\caption{Widths $\Delta$ corresponding to minima (lower branches) and
saddle points (upper branches) of the Ramakrishnan-Yussouf functional
restricted to Gaussian profiles for fcc and bcc solids. The asymptotic
linear behavior indicated by the dashed lines was calculated from
\eqpref{ELGasy}.}
\label{fig:biflogges}
\end{figure}

\begin{figure}
\caption{Free energies per particle of high density solids from the Ramakrishnan-Yussouff,
the two versions of fundamental-measure DFT, and from molecular
dynamics \protect\cite{Alder:68}. The de
Broglie wave length has been set to the particle diameter. The
asymptotic behavior indicated by the dotted lines is logarithmic in the free distance $a$ (see
\eqpref{Ftotasy}) in all cases.}
\label{fig:ftotlog}
\end{figure}

\begin{figure}
\caption{The space between two  nearest-neighbor sites (black dots) in
a crystal. The radii of  the  spheres is $\sigma/2\pm O(\Delta)$  where
$\Delta$ is the width of the (spherical) density peaks. The weighted densities in
region A  are  only influenced  from one  site  while in B  both sites
contribute. In  the remaining space  the excess  free energy densities
$\phi_i$ are neglible.}
\label{fig:regionab}
\end{figure}

\begin{figure}
\caption{Density profiles obtained from the fundamental-measure
theory FMT1.}
\label{fig:rhofmt}
\end{figure}

\begin{figure}
\caption{Density profiles obtained from the improved fundamental-measure
theory FMT2.}
\label{fig:rhofmtn}
\end{figure}

\begin{figure}
\caption{Illustration of the cells C for which the fundamental measure
theory predicts a constant density (a 2d analogon of the 3d crystal is
drawn). The circular arcs and their straight connection limit the set
of points whose distance to C is smaller than $\sigma/2$, i.e. the
region A. The point P cannot belong to C because otherwise P' would
belong to C' and some points in between would have distances smaller
than $\sigma/2$ from both C and C', i.e. region B would be nonempty.}
\label{fig:cell}
\end{figure}

\begin{figure}
\caption{Comparison of the density profiles from Monte-Carlo
simulation and the density-functional theory FMT2 for two bulk
densities. A Gaussian profile would correspond to a straight line in
this plot.}
\label{fig:rholog}
\end{figure}

\end{document}